\documentclass[showpacs, pra,twocolumn,preprintnumbers ,amsmath, amssymb, superscriptaddress, aps]{revtex4-2}
\usepackage{amsfonts}
\usepackage{color}
\usepackage{amsmath,amssymb}
\usepackage{pifont}
\usepackage{amssymb}  
\usepackage{bbold}
\usepackage{float}
\usepackage{subfloat}

\usepackage[caption=false]{subfig}
\usepackage{tikz}
\usepackage{makecell}
\usepackage{subfig}
\usepackage{pifont}   
\usepackage{graphicx} 
\graphicspath{{Figures5/}}
\usepackage{dcolumn}  
\usepackage{bm}       
\usepackage{multirow} 
\usepackage{placeins}
\usepackage[colorlinks]{hyperref}
\usepackage{mathtools}
\usepackage{appendix}

\def \be{\begin{align}}
	\def \ee{\end{align}}
\def \bea{\begin{eqnarray}}
	\def \eea{\end{eqnarray}}

\begin{document}

	\title{
		{Influence of Aharonov-Bohm flux and dual gaps on electron scattering in graphene quantum dots}}
	\date{\today}
	\author{Fatima Belokda}
	\affiliation{Laboratory of Theoretical Physics, Faculty of Sciences, Choua\"ib Doukkali University, PO Box 20, 24000 El Jadida, Morocco}
	\author{Ahmed Bouhlal}
	\affiliation{Laboratory of Theoretical Physics, Faculty of Sciences, Choua\"ib Doukkali University, PO Box 20, 24000 El Jadida, Morocco}
	\author{Ahmed Jellal}
	\email{a.jellal@ucd.ac.ma}
	\affiliation{Laboratory of Theoretical Physics, Faculty of Sciences, Choua\"ib Doukkali University, PO Box 20, 24000 El Jadida, Morocco}
	\affiliation{Canadian Quantum Research Center, 204-3002 32 Ave Vernon,  BC V1T 2L7, Canada}

	\pacs{}

	\begin{abstract}
		
.

We show how the Aharonov-Bohm flux (AB) \(\phi_i\) and the dual gaps $(\Delta_1, \Delta_2)$ can affect the electron scattering in graphene quantum dots (GQDs) of radius $r_0$ in the presence of an electrostatic potential \(V\). After obtaining the solutions of the energy spectrum, we explicitly determine the radial component of the reflected current $J_r^r$, the square modulus of the scattering coefficients $|c_m|^2$, and the scattering efficiency $Q$. Different scattering regimes are identified based on physical parameters such as incident energy \(E\), \(V\), $r_0$, dual gaps, and \(\phi_i\). In particular, we show that lower values of $E$ are associated with larger amplitudes of $Q$. Furthermore, it is found that $Q$ exhibits a damped oscillatory behavior with increasing the AB flux. In addition, increasing the external gap $\Delta_1$ resulted in higher values of $Q$. By increasing $\phi_i$, we show that the oscillations in $|c_m|^2$ disappear for larger values of $r_0$ and are replaced by prominent peaks at certain values of $E$ and angular momentum $m$. Finally, we show that  $J_{r}^r$ displays periodic oscillations of constant amplitude, which are affected by the AB flux.

		\vspace{0.25cm}
		\noindent PACS numbers: 73.22.Pr, 72.80.Vp, 73.63.-b\\
		\noindent Keywords: Graphene,  quantum dot, Aharonov-Bohm flux, dual gaps, electrostatic potential, scattering phenomenon.
		
	\end{abstract}
	\maketitle
\section{Introduction}

The creation of a band gap has spurred the development of various techniques to remotely control the conductance in graphene \cite{shimizu,omid}. One promising approach is to confine fermions within graphene by creating quantum dots (QDs), which limits the mobility of the electronic particles and allows better control over them. A variety of methods for achieving this confinement have been reported in the literature, as detailed in the review paper \cite{rozh}. These methods include the use of inhomogeneous magnetic fields \cite{martin,espino}, cylindrical potential symmetry \cite{chen}, and spatial modulation of the Dirac band gap \cite{giav}. In addition, confinement can be achieved by shaping the graphene flake into small nanostructures \cite{zebro,thom}, inducing a band gap through substrate interactions \cite{recher,pablo}, and using a co-precipitation technique \cite{saeed,kumar}. 
In order to confine Dirac fermions in graphene quantum dots, one must spatially limit their mobility inside an area of nanometer size, resulting in the formation of two-dimensional quantum structures. Atomic-scale control over electronic characteristics is made possible by this exact modification of the geometry and topology of graphene \cite{espino}, opening up a wide range of possible applications in photonics, electronics, and even spintronics  \cite{kats2006,cserti}. They might be used, for example, in the creation of extremely sensitive sensors \cite{wang}, ultrafast electronic devices, and quantum devices \cite{14} for quantum computing.

Electron diffusion in graphene quantum dots is influenced by several factors, such as the geometry of the quantum dot, the properties of the surrounding material, and external conditions such as temperature and electric field. These interactions can lead to interesting diffusion phenomena. Electron diffusion through graphene quantum dots can be modified by introducing energy gaps \cite{kim,njoum}, providing opportunities to manipulate electronic properties at the nanoscale. This ability to control electron diffusion opens up new avenues for the development of innovative electronic devices and advanced quantum technologies.

One of the fundamental phenomena of quantum interference is the Aharonov-Bohm (AB) effect, which directly depends on the phase coherence of charge carriers \cite{Aharonov1959}.
In graphene systems, the AB effect has been extensively studied, with particular emphasis on transport studies on nanorings and antidot lattices. In 2008, Russo {\it et al.} \cite{Russo2008} published some of the first experimental observations of this phenomenon in graphene. Indeed, they created rings of monolayer graphene and observed oscillations in the AB conductance caused by the constructive and destructive interference of electron waves orbiting the ring. 
The effect of AB on GQDs exposed to different external sources has been studied theoretically, see for example \cite{Bouhlal, Azar24}. Moreover, in \cite{Park2021}, the AB effect and quantum transport in bilayer graphene with magnetic quantum structures (rings and dots) were discussed.

On the other hand, \cite{26,27} reports that spatially variable and continuously changing mass terms can be realized through the use of boron nitride (BN) substrates. Using this result, it is possible to study the electrical properties of GQDs by demonstrating the presence of an inhomogeneous energy gap in the system. 
The potential to design inhomogeneous energy gaps in graphene-based devices is suggested by the ability to create spatially varying mass terms using BN substrates. The electronic behaviour of GQDs can be significantly influenced by this energy gap inhomogeneity, which may also allow the material properties to be manipulated and controlled. In light of the previous research, the current study aims to provide insight into how the presence of dual energy gaps can affect the electrical properties of graphene quantum dots (GQDs).

We study how the Aharonov-Bohm flux $\phi_i$ together with the dual gaps $(\Delta_1, \Delta_2)$ affect electron scattering in graphene quantum dots (GQDs) with radius $r_0$ under the influence of an electrostatic potential $V$. After deriving the energy spectrum solutions, we explicitly calculate the radial component of the reflected current $J_r^r$, the square modulus of the scattering coefficients $|c_m|^2$, and the scattering efficiency $Q$. Different scattering regimes are identified based on physical parameters such as the incident energy $E$, the potential $V$, the radius $r_0$, the dual gaps, and the AB flux $\phi_i$. In particular, lower values of $E$ correspond to larger amplitudes of $Q$, and $Q$ exhibits a damped oscillatory pattern with increasing flux. In addition, increasing the external gap $\Delta_1$ leads to higher $Q$ values. With increasing $\phi_i$ the oscillations in $|c_m|^2$ vanish for larger $r_0$ and are replaced by distinct peaks at specific $E$ values and angular momentum $m$. Finally, we observe that $J_r^r$ exhibits periodic oscillations of constant amplitude modulated by the flux.

\par The paper is organized as follows. In Sec.~\ref{model}, we introduce a theoretical model describing the system under consideration and determine the solutions of the energy spectrum. In Sec.~\ref{scattering}, by applying the boundary condition, we compute the analytical expressions for different quantities characterizing the scattering phenomenon. Sec.~\ref{DIS} presents the numerical analysis of the theoretical results under different configurations of the physical parameters. Finally, we conclude our work.

\section{Energy spectrum}\label{model}
\begin{figure}[ht]
	\centering
	\includegraphics[scale=0.35]{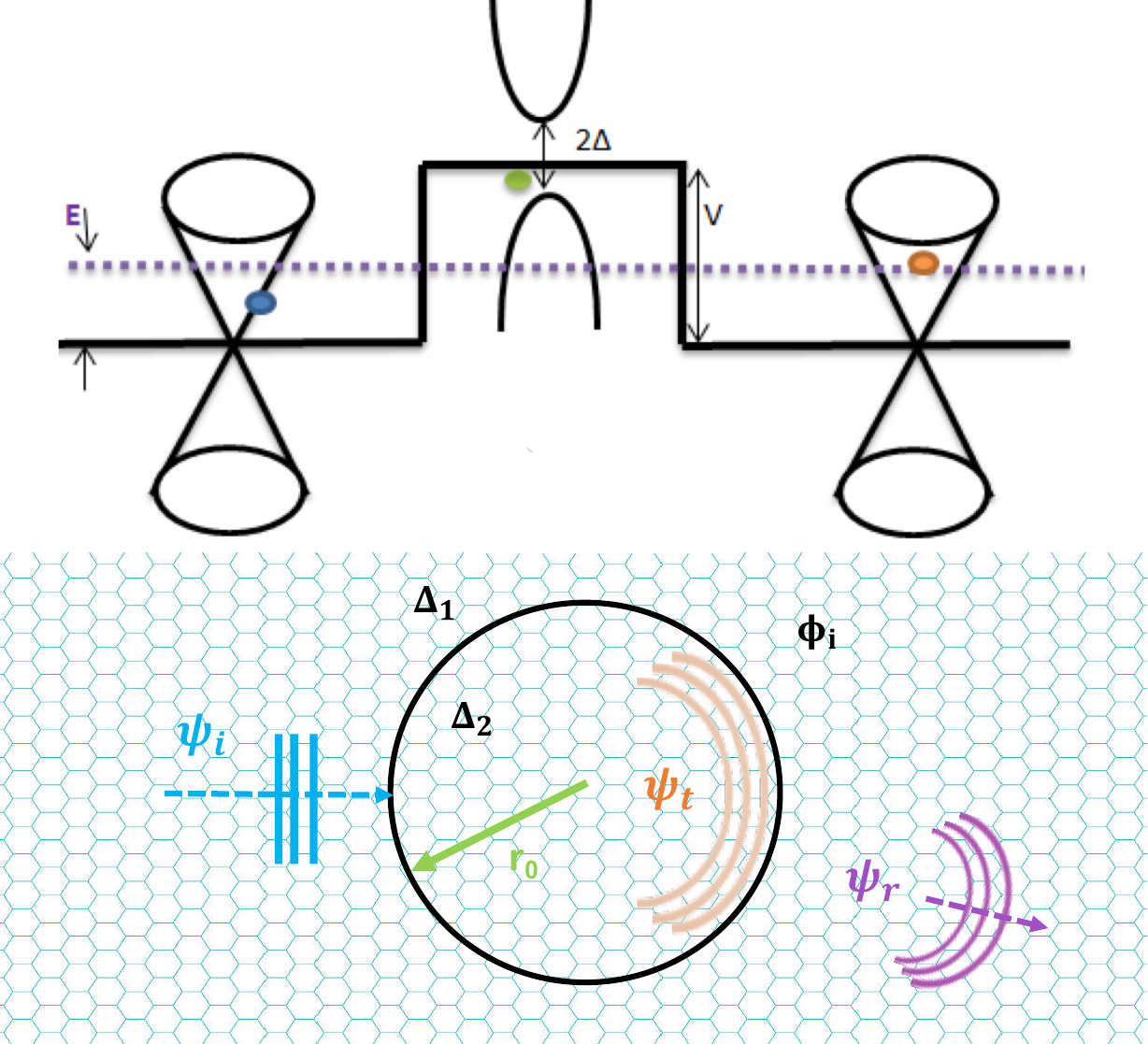}\\
	\caption{(color online) A graphene quantum dot with a radius $r_0$ subjected to Aharonov-Bohm flux  $\phi_i$, and an dual gaps $\Delta_j$.}\label{system}
\end{figure}

 As shown in Fig~\ref{system}, we consider GQDs of radius $r_0$ in the presence of AB flux $\phi_i$, electrostatic potential $V(r)$ and dual gaps $\Delta_j(r)$ ($j = 1, 2$). These external sources can be written as mathematically as
\begin{equation}
	V(r)=
	\left\{%
	\begin{array}{ll}
		0, & r>r_0 \\
		V ,&  r\leq r_0  
	\end{array}
	\right., \quad 	
	\Delta_j(r)=
	\left\{%
	\begin{array}{ll}
		\Delta_2 , &  r \leq r_0   \\ \Delta_1	, &  r>r_0
		
	\end{array}
	\right.		
\end{equation}
and the vector potential associated to Aharonov-Bohm flux is given by
\begin{equation}
	\vec{A}= \left\{%
	\begin{array}{ll}
		\frac{\phi_{AB}}{2 \pi r} \vec{e_\theta}, & r>r_0 \\
		0 ,&  r\leq r_0.  
	\end{array}
	\right.
\end{equation}
Note that in the forthcoming analysis, we use the rescalled flux $ \phi_i=\frac{\phi_{AB}}{\phi_0}$ where ${\phi_0}$ is the unit flux.

To characterize the movement of Dirac fermions within the honeycomb structure of carbon atoms, which are bonded by covalent bonds in graphene, under the influence of AB flux, we employ the following model represented by the Hamiltonian
\begin{equation}\label{hamilt}
	H_\tau=v_F\left(\vec{p} + e \vec{A}\right)\cdot\vec{\sigma}+V(r) \mathbb{I}_2 + \Delta_j(r) \sigma_z
\end{equation}
where $v_F=10^6 \mathrm{~m} / \mathrm{s}$ is the Fermi velocity, $\pi_i=p_i+e A_i$ are the conjugate momentum and $\sigma_i (i=x, y, z)$ are the Pauli matrices in the basis of the two sublattices of $A$ and $B$ atoms. 
The cylindrical symmetry of the present system dictates the use of polar coordinates $ (r,\theta)$. Now, by introducing the operator (in the unit $\hbar=\nu_{F}=1$)
\begin{equation}
	\partial_\pm= e^{\mp i\theta}
	\left(-i 
	{\partial_r}\pm\dfrac{1}{r} 
	{\partial\theta\pm i\frac{\phi_{i}}{r}}\right)
	\end{equation}
we map  \eqref{hamilt}
as follows
\begin{equation} \label{ham2}	
	H= \begin{pmatrix} V+\Delta_j &  \partial_+ \\  
		\partial_-	 & V-\Delta_j 
	\end{pmatrix}.
\end{equation}
In addition, it can be shown that the commutation relation $[H, J_z]=0 $ is fulfilled by the total momentum operator $J_{z}=L_{z}+\frac{1}{2}\sigma_{z}$. This provides the separability of the eigenspinors $\psi_{m}(r,\theta)$ of the Hamiltonian \eqref{ham2} into radial $R^{A,B}(r) $ and angular $\chi^{A,B}(\theta)$ parts, such as
\begin{equation}   
	\psi_m{(r,\theta)}=
	\begin{pmatrix} R^{A}_{m}(r)\chi^{A}(\theta)  \\R^{B}_{m+1}(r)\chi^{B}(\theta) \label{55}
	\end{pmatrix} 
\end{equation}
where the eigenstates of $J_z$ are
\begin{equation}
	\chi^{A}(\theta)=\dfrac{e^{im\theta}}{\sqrt{2\pi}}\begin{pmatrix} 1\\  0  \end{pmatrix}, \qquad 
	\chi^{B}(\theta)=\dfrac{e^{i(m	+1)\theta}}{\sqrt{2\pi}}\begin{pmatrix} 0\\  1  \end{pmatrix}
\end{equation}
and $ m=0,\pm1,\pm2, \cdots $ being the angular momentum quantum number.

By finding the radial components we complete the eigenspinor derivation and obtain the energy spectrum solutions. To do this we solve the eigenvalue equation $H\psi_m{(r,\theta)}=E\psi_m{(r,\theta)}$ in the regions outside $ r>r_0$ and inside $ r\leq r_0$ of the GQDs (see Fig.~\ref{system}). As a result, we show that the radial components $ R^{A}_{m}(r)$ and $ R^{B}_{m+1}(r)$ actually satisfy two coupled differential equations for $ r>r_0 $
\begin{align}
	& \left(-i{\partial_r} +i\frac{m+\phi_i}{r}\right) R^{A}_{m}=(E+\Delta_1) R^{B}_{m+1} \label{77}
	\\
	&
	\left(-i{\partial_r} -i\frac{m+\phi_i+1}{r}\right) R^{B}_{m+1}=(E-\Delta_1) R^{A}_{m} \label{88}.
\end{align}
By injecting \eqref{77} into \eqref{88}, we find 
a second differential equation
\begin{equation}\label{Bessel}
	\left(r^{2}{\partial_r^{2}}+ r{\partial_r}+r^{2}k_1^{2}-(m+\phi_i)^{2}\right) R^{A}_{m}(r)=0
\end{equation}
where we have set the wave vector
\begin{align}
	k_1=\sqrt{{E^{2}}-\Delta_1^{2}}.	
\end{align}
We show that the Bessel functions $ J_m(k_1r) $ is the solution of \eqref{Bessel}. 
Usually, for a normalized incident plane 
\begin{align}	
 \psi=\dfrac{1}{\sqrt{2}} e^{ikx} \begin{pmatrix} 1\\  1 \label{1010}\end{pmatrix}	
\end{align}
with $x=r\cos\theta$
	we can map it in terms of  $ J_m$ as 
\begin{equation}
	\psi
	=\dfrac{1}{\sqrt{2}}\sum_m{i^{m}}J_m(k_1r)e^{im\theta}\begin{pmatrix} 1\\  1 \label{1010}\end{pmatrix}.
\end{equation}
Applying this to our solutions for region $ r> r_0$, 
we end up with
the incident and reflected spinors 
\begin{align}
	\label{inc}		\psi_\text{inc}{(r,\theta)}&=\sqrt{\pi}\sum_{m}{i^{m+1}}\\	&\left[-iJ_{m}(k_1r) \chi^A(\theta) +\mu_1 J_{m}(k_1r)
	\chi^B(\theta)	\right]\nonumber\\
	\label{ref}
	\psi_\text{ref}{(r,\theta)}&=\sqrt{\pi}\sum_{m}{i^{m+1}}a_{m+\phi_i}\\	&\left[-iH^{(1)}_{m+\phi_i}(k_1r)\chi^A(\theta)+  \mu_1H^{(1)}_{m+\phi_i+1}(k_1r)
	\chi^B(\theta)	\right]	\nonumber
\end{align}	
where $H^{(1)}_{m}(k_1r) $ is the Hankel function of the first kind, $a_{m+\phi_i} $ are the scattering coefficients. Here  dimensionless parameters 
\begin{align}
	\mu_1=\sqrt{\frac{E-\Delta_1}{E+\Delta_1}}
\end{align}
is defined
in terms of the first energy gap $\Delta_{1}$, which becomes one for 
$\Delta_{1}=0$.

As for the second region $ r\leq r_0$, which includes the potential $V$ and the second energy gap $\Delta_{2}$, the eigenvalue equation allows us to get 
\begin{align}
	& \left(-i{\partial_r} +i\frac{m}{r}\right) R^{A}_{m}=(E-V+\Delta_2) R^{B}_{m+1} \label{1313}
	\\
	&
	\left(	-i{\partial_r} -i\frac{m+1}{r}\right) R^{B}_{m+1}=(E-V-\Delta_2) R^{A}_{m}\label{1414}.
\end{align}
Using the same process as  before, we  end up with a second differential equation for $R^{A}_{m}$
\begin{align}\label{trss}
	(r^{2}{\partial_r^{2}}+ r{\partial_r}+r^{2}k_2^{2}-(m)^{2}) R^{A}_{m}=0
\end{align}
where the associated wave vector is given by
\begin{align}
	k_2=\sqrt{(E-V)^{2}-\Delta_2^{2}}.
\end{align}
From \eqref{trss} we derive the transmitted spinor solution
\begin{align}\label{tra}
	\psi_\text{tra}{(r,\theta)}&=\sqrt{\pi}\sum_{m}{i^{m+1}}b_{m}\\	&\left[-iJ_{m}(k_2 r)\chi^A(\theta)
	+\mu_2 J_{m+1}(k_2 r) \chi^B(\theta)
	\right]\nonumber
\end{align}
where $ b_{m }$ are the scattering coefficients
and the parameter $\mu_2$ is given by 
\begin{align}
	\mu_2 =\sqrt{\frac{E-V-\Delta_2}{E-V+\Delta_2}}.
\end{align}
In the next step, we will examine how the previous results can be applied to analyze Dirac electron scattering in the present system.

\section{Scattering phenomenon}\label{scattering}

To study the scattering problem, we should first determine the scattering coefficients $ a_{m}$ and $ b_{m}$. For this we use the boundary condition at the interface $ r=r_0$, which is satisfied by the spinors (\ref{inc}-\ref{ref}) and \eqref{tra}. Otherwise, we have  
\begin{align}
	\psi_\text{inc}(r_0,\theta)+\psi_\text{ref}{(r_0,\theta)}=\psi_\text{tra}{(r_0,\theta)}. 
\end{align}
After substitution, we establish two relations between Bessel and Hankel functions. They are 
\begin{align}
	&	J_{m}(k_1 r_0)+a_{m+\phi_i}H^{(1)}_{m + \phi_i}(k_1 r_0)=b_{m}J_{m}	(k_2 r_0)	
	\\
	&
	\mu_1J_{m+1}(k_1 r_0)+\mu_1a_{m+\phi_i }H^{(1)}_{m + \phi_i +1}	(k_1 r_0)\nonumber\\
	&= \mu_2 b_{m}J_{m+1}(k_2 r_0)		
\end{align}
which can be solved to obtain the scattering coefficients
\begin{align}
	&a_{m+ \phi_i}=\\
	&\frac{{ \mu_2J_{m}(k_1 r_0)J_{m+1}	(k_2 r_0) -\mu_1J_{m}(k_2 r_0)J_{m +1}(k_1 r_0)}}{\mu_1 J_{m}(k_2 r_0)H^{(1)}_{m+\phi_i+1}	(k_1 r_0)-\mu_2 J_{m+1}	(k_2 r_0)H^{(1)}_{m+\phi_i}	(k_1 r_0)}	\nonumber			
	\\
	&
	b_{m}=\\
	&\frac{{\mu_1J_{m}(k_1 r_0)H^{(1)}_{m+\phi_i+1}(k_1 r_0)- \mu_1J_{m+1}	(k_1 r_0)H^{(1)}_{m+\phi_i}(k_1 r_0) }}{\mu_1J_{m}	(k_2 r_0)H^{(1)}_{m+\phi_i+1}	(k_1 r_0)-\mu_2 J_{m+1}	(k_2 r_0)H^{(1)}_{m+\phi_i}	(k_1 r_0)}	\nonumber
\end{align} 
	
At this level we determine the radial current associated with the present system. Then, we use the Hamiltonian \eqref{hamilt} to obtain the current density 
\begin{equation}
	\vec{j}=\psi^{\dagger}\vec{\sigma}\psi  
\end{equation}
such that $\psi=\psi_\text{tra}$  inside  and  $ \psi=\psi_\text{inc}+\psi_\text{ref}$
 outside the GQDs. Using its projection on the unit vector $\vec{e}_r$, i.e., $j_{r}=\vec{j} \cdot \vec{e}_r$, we derive the radial component as
\begin{equation} 
	j_r=\psi^{\dagger} \begin{pmatrix} 0& \cos{\theta}-i\sin{\theta} \\ \cos{\theta}+i\sin{\theta}&0 
	\end{pmatrix}\psi.
\end{equation} 
As far as the reflected spinor \eqref{ref} is concerned, we get the reflected radial current
\begin{align} \label{jrr}
	j^{r}_r=\dfrac{1}{2}\sum_{m=0}^{\infty} A_{m}(k_1r)\begin{pmatrix} 0& {e^{-i\theta}} \\e^{i\theta}& 0 \end{pmatrix}\sum_{m=0}^{\infty} B_{m}(k_1r)
\end{align} 
where we have defined the two functions 
\begin{widetext}
	\begin{align} \label{311}
		&A_{m} 
		=(-i)^{m+1}\left[iH^{(1)^{*}}_{m+\phi_i}(k_1r)
		\begin{pmatrix}a^{*}_{m}e^{-im\theta}& a^{*}_{-(m+1)}  e^{+im\theta}\end{pmatrix}+\mu_1H^{(1)^{*}}_{(m+\phi_i+1)}(k_1r)\begin{pmatrix}a^{*}_{-(m+1)}e^{-i(m+1)\theta}& a^{*}_{(m)}e^{i(m+1)\theta}\end{pmatrix}\right]
		\\
		& B_{m} 
		=(-i)^{m+1}\left[iH^{(1)}_{m+\phi_i}(k_1r)\begin{pmatrix}a_{m}e^{im\theta}\\a_{-(m+1)}e^{-im\theta}\end{pmatrix}+\mu_1H^{(1)}_{(m+\phi_i+1)}(k_1r)\begin{pmatrix}a_{-(m+1)}e^{-i(m+1)\theta}\\a_{(m)}
			e^{i(m+1)\theta}\end{pmatrix}\right]\label{322}.
	\end{align} 
	\end{widetext}
To simplify the above relation and to give a better understanding, let us consider the asymptotic behavior of the Hankel function for large arguments, i.e., $k_1r \gg 1$. It is given by
\begin{equation} 
	H_{m}(k_1r)\simeq\sqrt{\dfrac{2}{\pi k_1 r}}e^{i\left(k_1 r-\frac{(2m+1)\pi}{4}\right)}\label{2626}
\end{equation}
which can be substituted into (\ref{311}-\ref{322})   
to approximate the reflected radial component \eqref{jrr} as follows
\begin{equation}
	j^{r}_r=\frac{2}{\pi k_1 r}\sum_{m=0}^{\infty}|c_m(\phi_i)|^{2} \left(\cos[(2m+1)\theta]\left(1+\mu_1^{2}\right)+2\mu_1\right)\label{jrra}
\end{equation} 
where the square modulus of the scattering coefficients $	|c_m(\phi_i)|^2$ are expressed in terms of
\begin{align}
	|c_m(\phi_i)|^{2}=\frac{1}{2}\left(|a_{m+\phi_i}|^{2}+|a_{-(m+\phi_i+1)}|^{2}\right).	
\end{align}

At this stage, we examine other important quantities related to the scattering phenomenon. In particular, we can use \eqref{2626} in the limit \( k_1r \rightarrow \infty \) to calculate the scattering cross section \( \sigma \), which is defined by
\begin{equation}
	\sigma=\frac{I^{r}_r}{(I_i/A_u)}	
\end{equation}
where the total reflected flux $I^{r}_r $ can be calculated by integrating the reflected radial current over a concentric circle
\begin{align}
	I^{r}_r=\int^{2\pi}_{0}j^{r}_r(\theta)r d\theta
\end{align}
to get a relation in terms of the square modulus of the scattering coefficients
\begin{equation} 
	I^{r}_r	
	=\frac{8}{E+ \Delta_{1}}\sum_{m=0}^{\infty}|c_m(\phi_i)|^{2}\label{3030}. \end{equation}
It is important to note that $I^{r}_r$ depends on $\Delta_{1}$, which will be used extensively in numerical analysis. Furthermore, it is obvious to obtain the incident flux per unit area $I_i/A_u =1$ using the incident spinor \eqref{1010}. 
The scattering cross section $\sigma$ is therefore equal to the total reflected flux $I^{r}_r$. 
 
 To compare the scattering on the GQDs of different sizes, it is convenient to introduce the scattering efficiency $Q$. It is defined as the ratio between $\sigma$ and the geometric cross-section.
\begin{align}
	Q=\frac{\sigma}{2r_0}	
\end{align}
which gives
\begin{equation}
	Q=\frac{4}{r_0(E+\Delta_{1})}\sum_{m=0}^{\infty}|c_m(\phi_{i})|^{2}\label{3030}.
\end{equation}
We emphasize that the results obtained should be reduced to those in \cite{Jellal18} as long as one takes $\Delta_{1}=0$. We will analyze our results numerically and discuss the main features of the present system. It will also be noted that \eqref{jrra} depends strongly on the energy gap $\Delta_{1}$, which will certainly play a crucial role in the forthcoming analysis and makes a difference with respect to the analysis reported in \cite{Jellal18}.

\section{Discussions}\label{DIS}

\begin{figure*}[ht]
	\centering
	\includegraphics[scale=0.4]{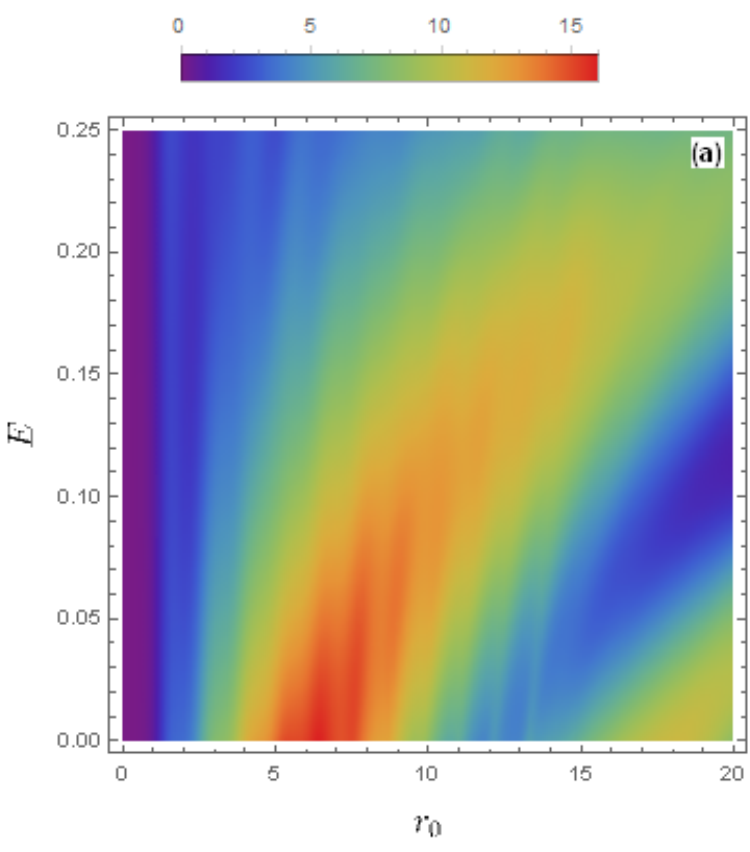}	
	\includegraphics[scale=0.4]{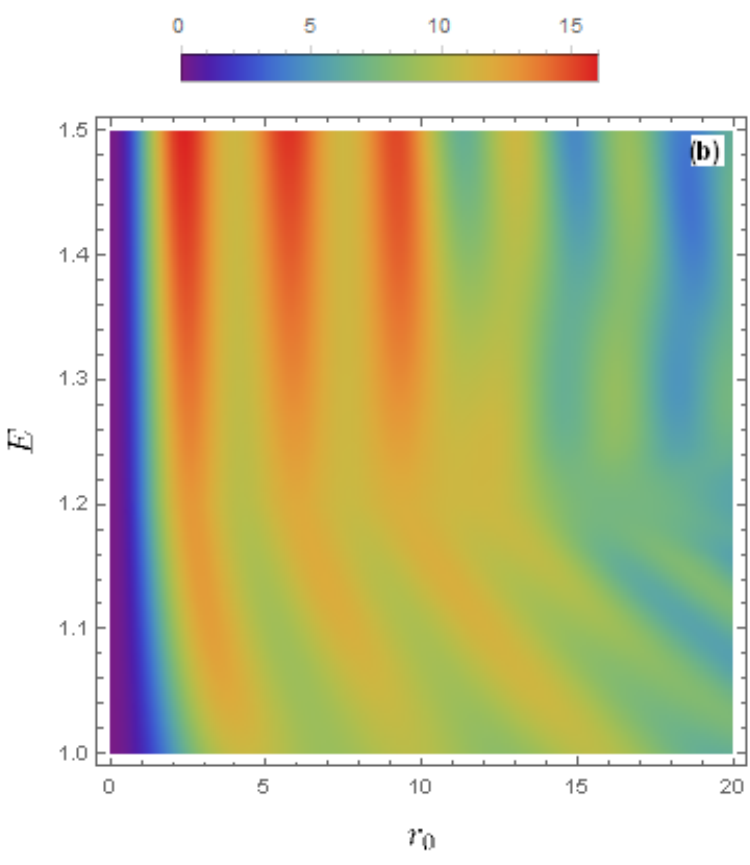}
	\includegraphics[scale=0.4]{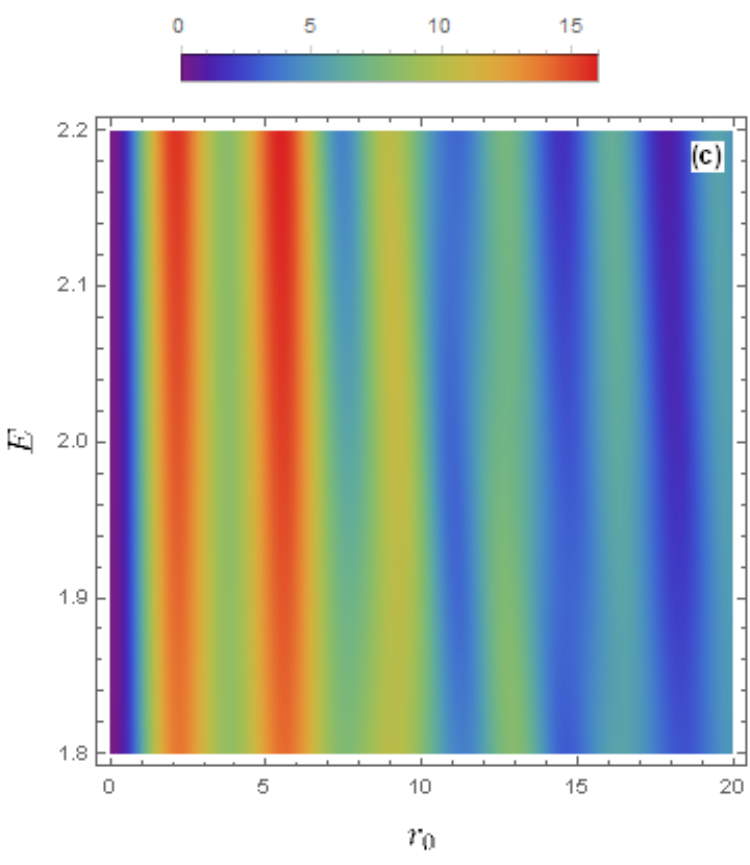}\\
	\includegraphics[scale=0.4]{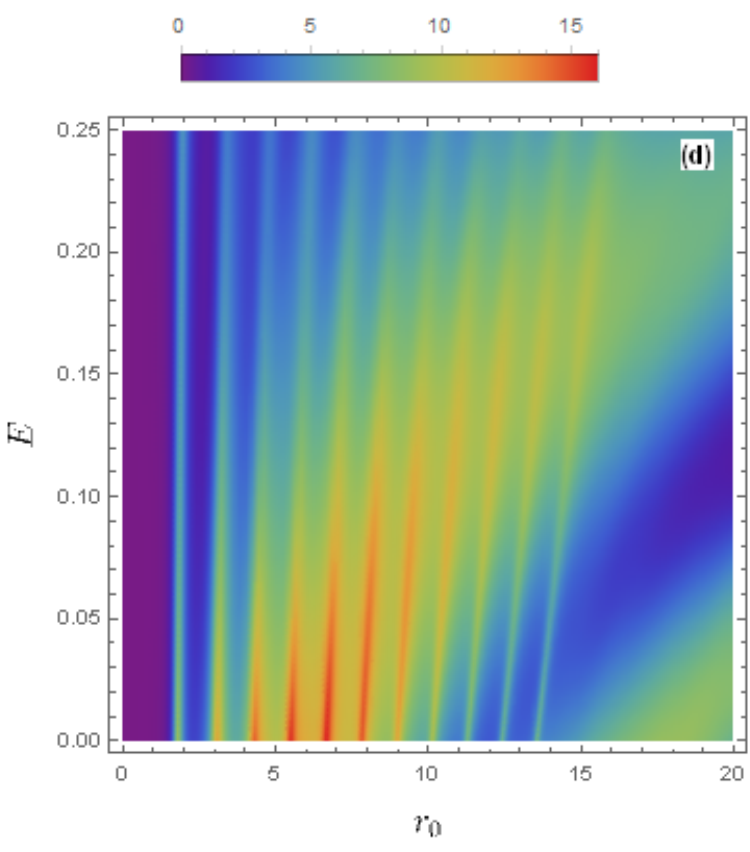}
	\includegraphics[scale=0.4]{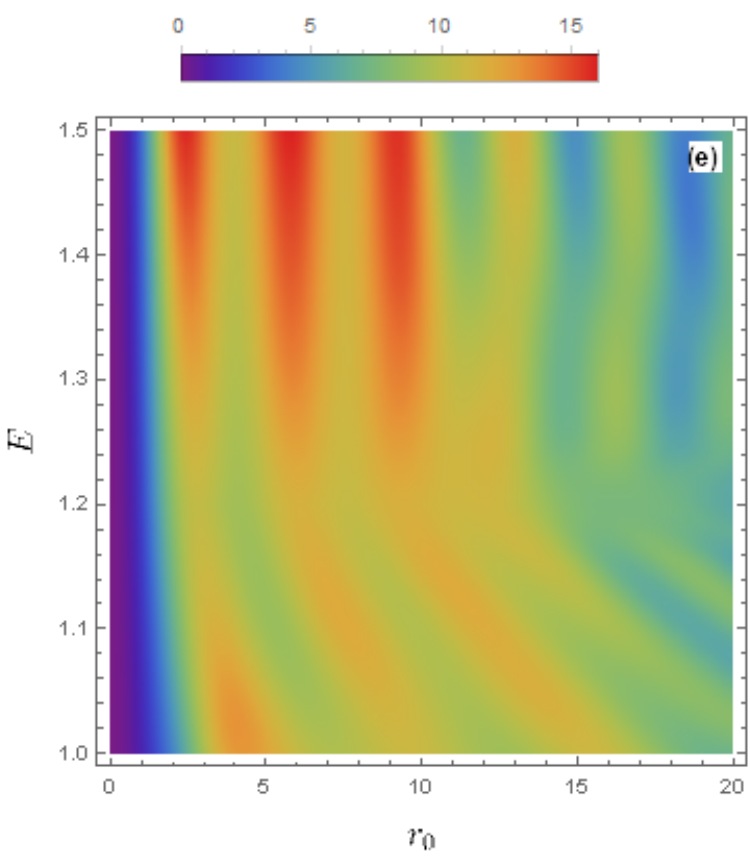}
	\includegraphics[scale=0.4]{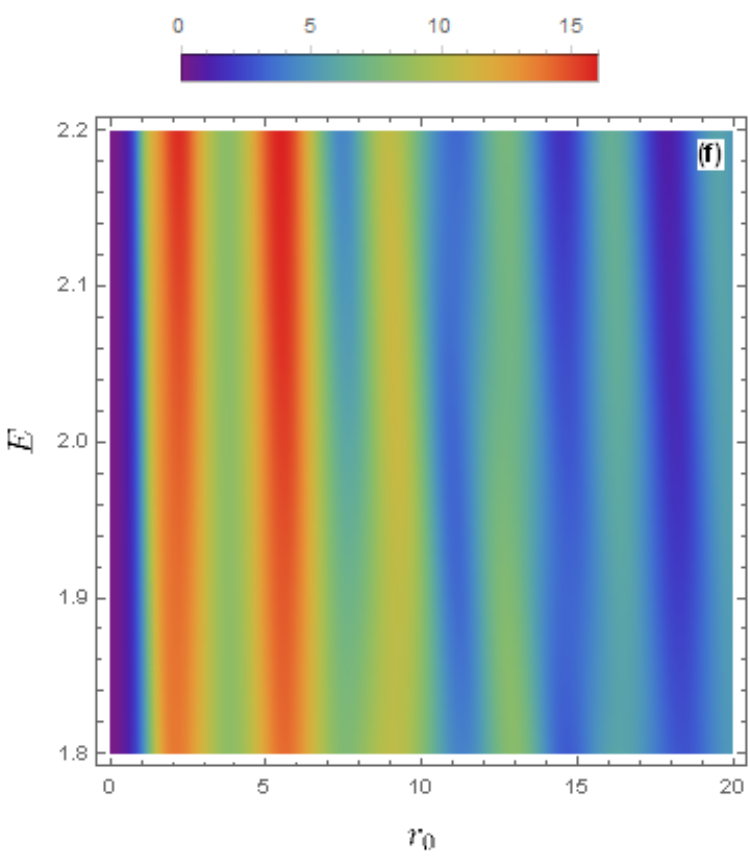}\\
	\includegraphics[scale=0.4]{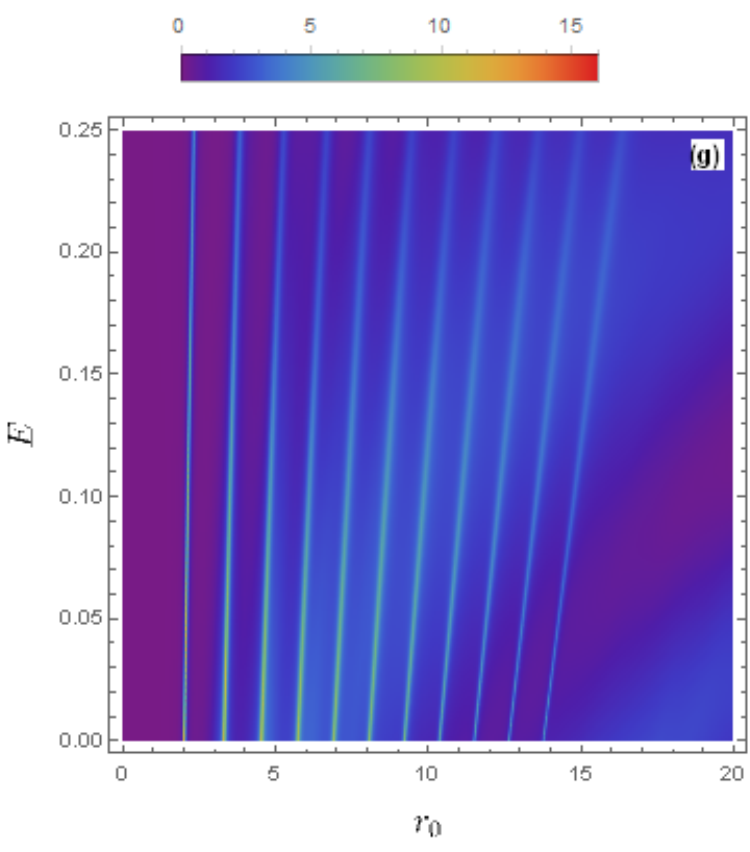}
	\includegraphics[scale=0.4]{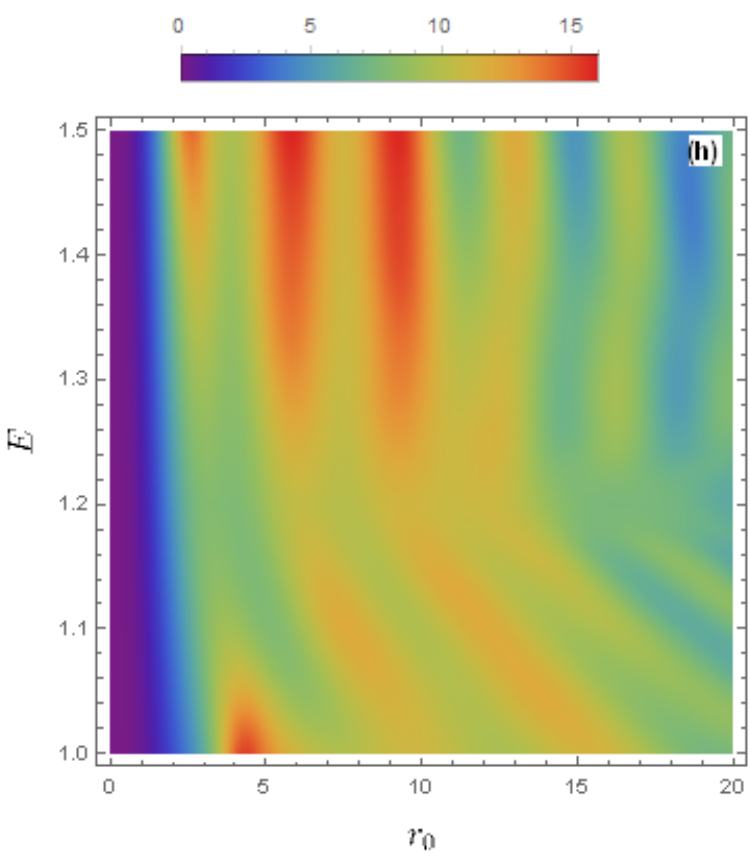}
	\includegraphics[scale=0.4]{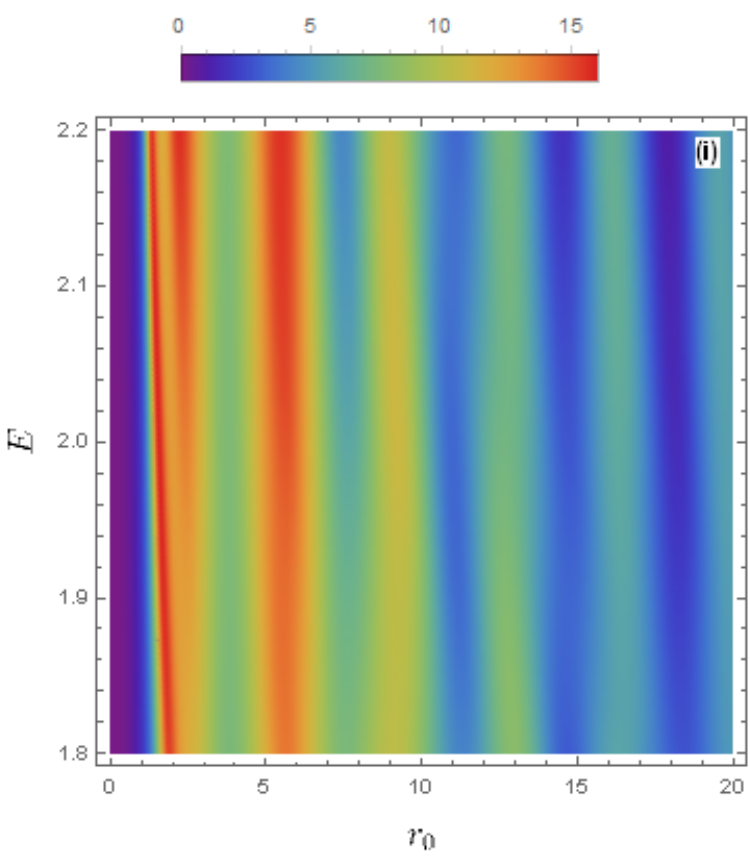}\\
	\caption  {(color online) Density plot of the scattering efficiency $Q$ as a function of the radius $r_0$  and
		the incident energy $E$ for 	
		(a,b,c): $\phi_i$=0.5, (d,e,f): $\phi_i$= 1.5 and (g,h,i): $\phi_i$=2.5,  first column: $E<V_-$, second column: $<V_- E< V_+$, and third column: $E > V_+ $. } \label{f2}
\end{figure*}

Figure \ref{f2} shows a density plot of the scattering efficiency $Q$ as a function of radius $r_0$ and incident energy $E$. In the first region, where $E < V_-$, the bands appear vertical, corresponding to $m+1$ modes, with large values observed for $5 < r_0 < 10$. As the AB flux $\phi_i$ increases, $Q$ decreases. In the second region, $V_- < E < V_+$, $Q$ forms broad and narrow vertical bands at high energies, while the bands are tilted at lower energies. In the third region, where $E > V_+$, $Q$ varies only in broad and narrow vertical bands, decreasing with increasing $r_0$. Notably, the effect of increasing $\phi_i$ is significant in all regions except the first, where $E < V_-$.

\begin{figure*}[ht]
	\centering
	\includegraphics[scale=0.4]{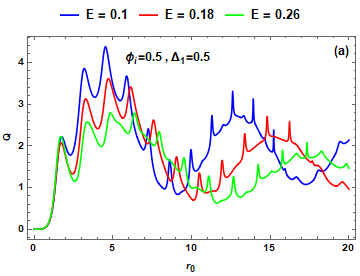}	
	\includegraphics[scale=0.41]{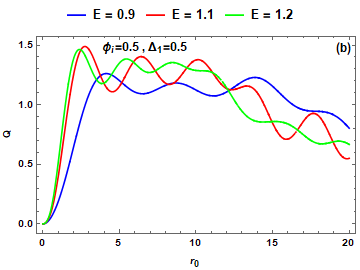}
	\includegraphics[scale=0.41]{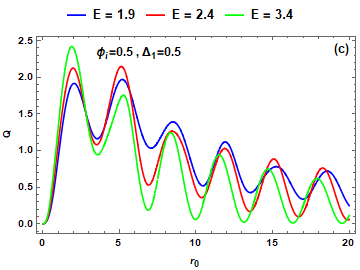}\\
	\includegraphics[scale=0.41]{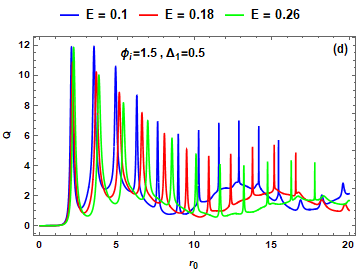}
	\includegraphics[scale=0.41]{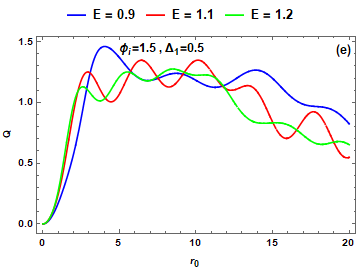}
	\includegraphics[scale=0.41]{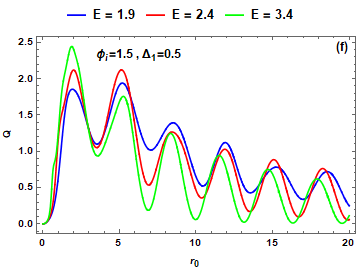}\\
	\includegraphics[scale=0.41]{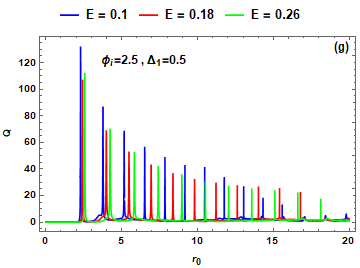}
	\includegraphics[scale=0.41]{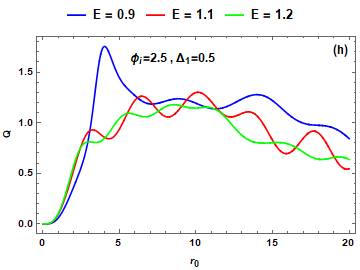}
	\includegraphics[scale=0.41]{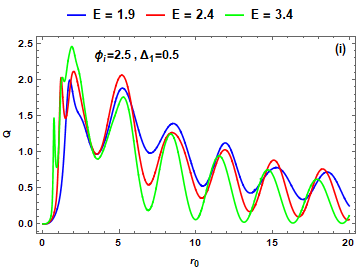}\\
	\caption  {(color online) The scattering efficiency $Q$ as a function of the radius $r_0$  for $V = 1$,  $\Delta_2 = 0.2$,  $\Delta_1 = 0.5$,  different values of the incident energy $E$, and $\phi_i=0.5,1.5,2.5$ . Three regions are labeled as (a,d,g): $E<V_-$, (b,e,h): $V_-< E< V_+$, and (c,f,i): $V_+ < E$.} \label{f3}
\end{figure*}

In Fig.~\ref{f3} we plot the scattering efficiency $Q$ as a function of the radius $r_{0}$ for different choices of $(E,V,\Delta_{1},\Delta_{2},\phi_{i})$. We notice that $Q$ varies differently according to three regimes depending on the incident energy $E$. The energy range $E<V_-$ is shown in Fig. \ref{f3}(a,d,g). For $\phi_{i}$=0.5 in Fig. \ref{f3}a, we observe that $Q$ varies linearly for small values of $r_{0}$, but as $r_0$ increases, $Q$ varies in the form of oscillations with different amplitudes. Increasing the  AB flux to $\phi_{i}$=1.5 in Fig. \ref{f3}d, we see that $Q$ increases to reach maximum peaks at the value $Q$=12, and these peaks diminish as $r_{0}$ increases. In Fig. \ref{f3}g for $\phi_{i}$=2.5, $Q$ reaches maximum peaks at $Q$=120, and as $r_{0}$ increases, these peaks cancel out.
The energy range $V_-<E<V_+$ is plotted in Fig. \ref{f3}(b,e,h). We observe that $Q$ varies linearly for small values of $r_{0}$. As we increase $r_{0}$, we see that $Q$ oscillates around approximate values and depends on the incident energy $E$. As the AB flux increases to $\phi_{i}$=1.5 in Fig. \ref{f3}e and Fig. \ref{f3}h, $Q$ varies in the same way. 
Fig. \ref{f3}(c,f,i) is associated with the energy range $E>V_+$. It appears that $Q$ varies linearly for small values of $r_{0}$. As $r_{0}$ increases, $Q$ begins to oscillate with damped oscillations for all values of  $\phi_{i}$.

\begin{figure*}[ht]
	\centering
	\includegraphics[scale=0.4]{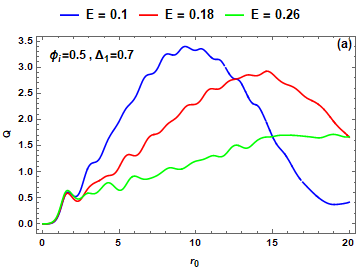}	
	\includegraphics[scale=0.4]{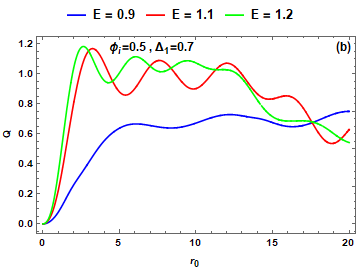}
	\includegraphics[scale=0.4]{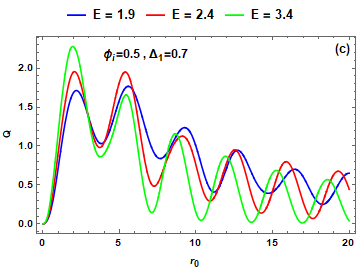}\\
	\includegraphics[scale=0.4]{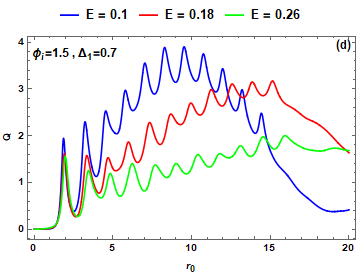}
	\includegraphics[scale=0.4]{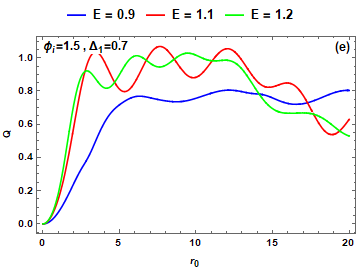}
	\includegraphics[scale=0.4]{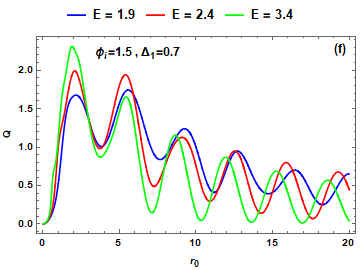}\\
	\includegraphics[scale=0.4]{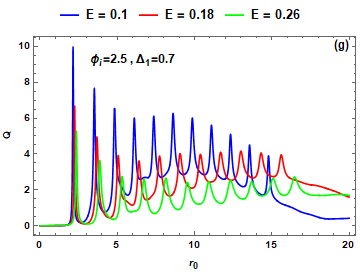}
	\includegraphics[scale=0.4]{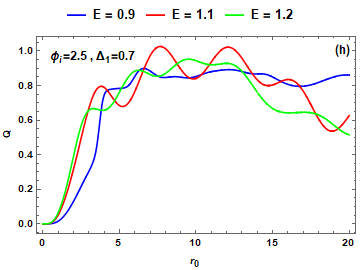}
	\includegraphics[scale=0.4]{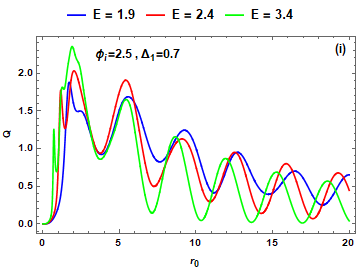}\\
	\caption  {(color online) The same as Fig. \ref{f4} but $\Delta_1$ increased to = 0.7. }
\label{f4}
\end{figure*}

The same as in Fig.~\ref{f3}, but we increase the energy gap to $\Delta_{1}$=0.7 to emphasize its effect on $Q$ in Fig.\ref{f4}.
Indeed, the first range $E<V_-$ is plotted in Fig. \ref{f4}(a,d,g). Taking $\phi_{i}$ = 0.5 in Fig. \ref{f4}a, we see that $Q$ varies linearly when $r_{0} < 2$, then it increases according to the values of $E$ until it reaches a maximum value $Q$ = 3, corresponding to $r_{0}$ = 10, and then it starts to decrease. 
We note that $Q$ fluctuates with small oscillations and increases until $r_{0}$ = 10, at which point it begins to decrease when we increase the  AB flux for $\phi_{i}$ = 1.5 in Fig. \ref{f4}d. $Q$ fluctuates oscillatorily in Fig. \ref{f4}g with $\phi_{i}$ = 2.5, showing the appearance of peaks at $r_{0}$ = 2. It then begins to decline as $r_{0}$ increases.
For $V_- < E < V_+$ in Fig. \ref{f4}(b,e,h), $Q$ varies linearly for small values of $r_{0}$.  As long as $r_{0}$ increases, $Q$ shows small oscillations for all values of the incident energy $E$ and $\phi_{i}$.
In Fig. \ref{f4}(c,f,i) associated with $E > V_+$, $Q$ increases linearly until it reaches a maximum $Q$ = 2 for $r_{0}< 2$. As $r_{0}$ increases, $Q$ begins to oscillate with damped oscillations for all energy values. 
We note that the increase in  AB flux affects $Q$ for the regime $E<V_-$, and for the other regimes the results remain similar to those obtained in \cite{belokda}.

\begin{figure}[ht]
	\centering
	\includegraphics[scale=0.335]{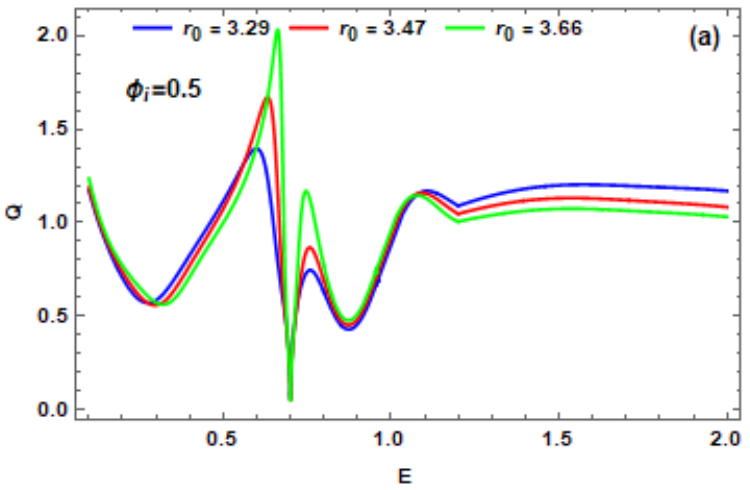}	
	\includegraphics[scale=0.335]{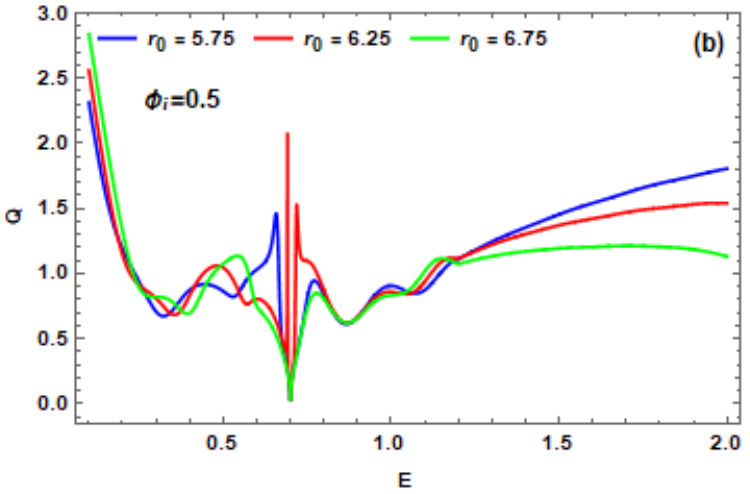}\\
	\includegraphics[scale=0.335]{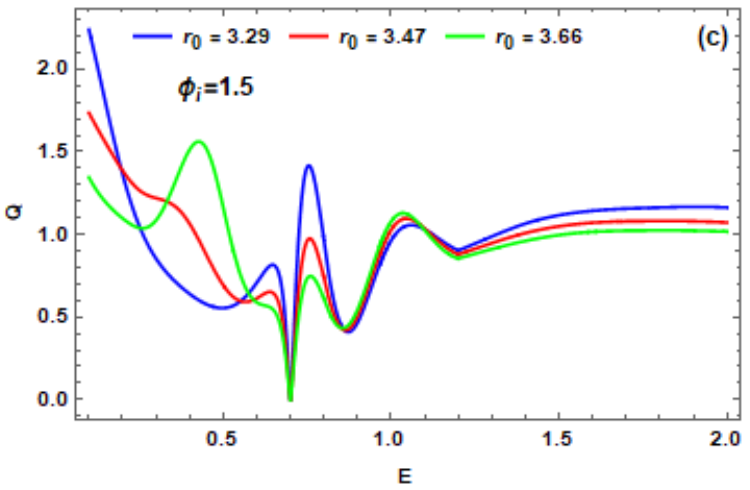}
	\includegraphics[scale=0.335]{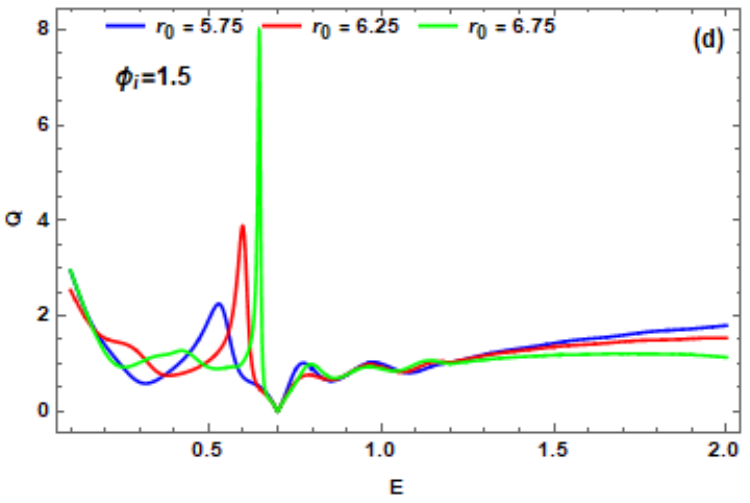}\\
	\caption  {(color online) The scattering efficiency $Q$ as function of the incident energy $E$ for  $\Delta_2=0.2$, $\Delta_1=0.7$ and $ V = 1$, with (a,c): $r_0$ = 3.29, 3.47, 3.66, (b,d): $r_0$ = 5.75,6.25, 6.75, (a,b): $\phi_{i}=0.5$ and (c,d):  $\phi_i=1.5$} \label{f5}
\end{figure}

For GQDs of different sizes $r_{0}$, the variation of $Q$ as a function of the incident energy $E$ is plotted in Fig. \ref{f5}. Note that $Q$ is not zero for $E=0$ and decreases to a minimum for $\phi_{i}$=0.5 in Fig. \ref{f5}a. In the case of $E=\Delta_{1}$=0.7, $Q$ reaches a minimum at zero, and as $E$ rises, $Q$ falls rapidly to a maximum of $Q$=2. $Q$ fluctuates a bit as $E$ rises before becoming constant. In Fig. \ref{f5}c, $Q$ falls for $\phi_{i}$=1.5 and reaches a minimum for $E$=$\Delta_{1}$=0.7, after which it oscillates again. As long as $E$ increases, $Q$ approaches stability. 
We further increase the size of the GQDs in Fig. \ref{f5}(b,d). Fig. \ref{f5}b illustrates this for $\phi_{i}$=0.5, where $Q$ takes a maximum value of 2.8 at $E$=0. Increasing $E$ causes $Q$ to fall until $E$=0.7=$\Delta_{1}$, at which point $Q$ begins to fluctuate and then goes through a linear variation to a constant.
We increase the  AB flux to $\phi_{i}$=1.5 in Fig. \ref{f5}d. For small values of $E$, $Q$ decreases, but when $E$=0.7=$\Delta_{1}$, $Q$ takes maximum peaks, reaching $Q$=8 for $r_{0}$=6.25. As $E$ increases, $Q$ shows small oscillations and then becomes linearly constant. 
 In summary, we find that increasing the  AB  flux $\phi_{i}$ suppresses the maximum values of $Q$ for $E <$ $\Delta_{1}$ for small sizes $r_{0}$, while for large sizes of $r_{0}$ increasing  $\phi_{i}$ gives maximum peaks of $Q$, especially for $r_{0}$=6.25.

\begin{figure}[ht]
	\centering
	\includegraphics[scale=0.34]{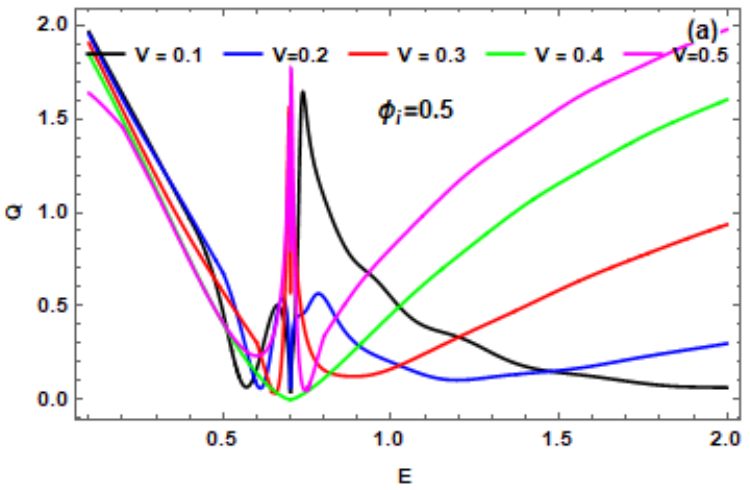}	
	\includegraphics[scale=0.33]{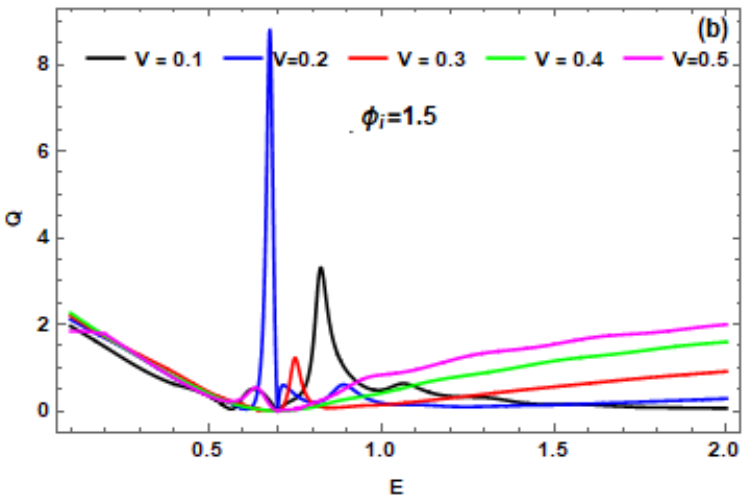}
	\caption{(color online) The same as Fig.~\ref{f5} but with potential values
		$V=0.1, 0.2, 0.3, 0.4, 0.5$, $\Delta_2$=0.3 and $r_0$ = 4.}
	\label{f6}
\end{figure}

Fig. \ref{f6} shows the scattering efficiency $Q$ as a function of the incident energy $E$ but with different values of the potential $V$. For $\phi_{i}$=0.5 in Fig. \ref{f6}a, $Q$ reaches a maximum at $E$=0, and as $E$ increases, $Q$ decreases linearly until it reaches $0$. For $E=$ $\Delta_{1}$=0.7, $Q$ reaches a maximum, while for $E>\Delta{1}$, $Q$ increases linearly for $V$=0.5, 0.4, 0.3, but for $V$=0.2, 0.1, $Q$ decreases and then remains almost stable. In Fig. \ref{f6}b for $\phi_{i}$=1.5, $Q$ varies in the same way, but when $E$=$\Delta_{1}$=0.7, $Q$ tends to reach maximum peaks, especially for $V$=0.2 where $Q$=8.

\begin{figure}[ht]
	\centering
	\includegraphics[scale=0.335]{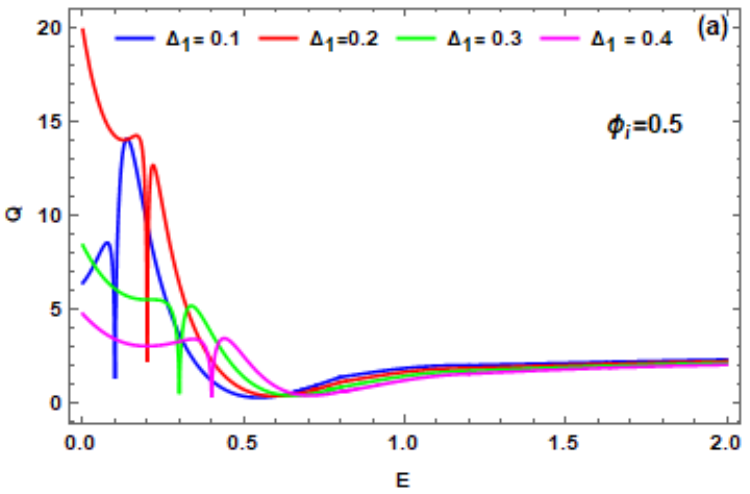}	
	\includegraphics[scale=0.335]{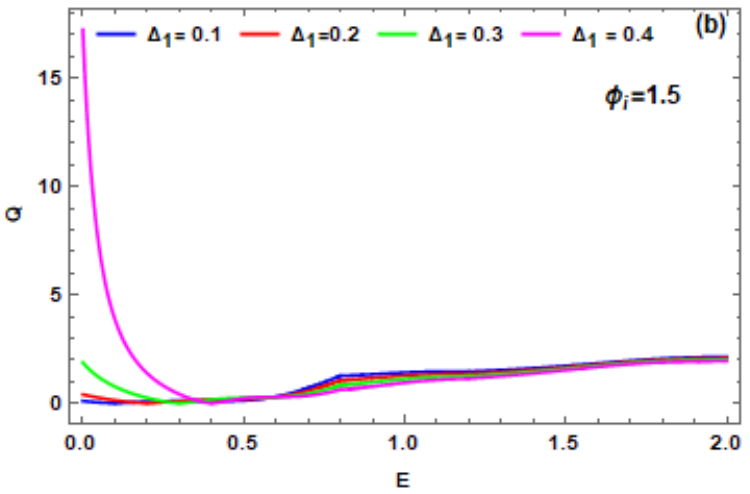}
	\caption{(color online) The same as Fig. \ref{f5} but with outside gap values
		$\Delta_1 = 0.1, 0.2, 0.3, 0.4$  and $r_0$ = 2.}
		\label{f7}
\end{figure}

In Fig. \ref{f7}, we plot the scattering efficiency $Q$ as a function of the incident energy $E$ for different values of the energy gap $\Delta_{1}$. In Fig. \ref{f7}a for $\phi_{i}$ = 0.5, we observe that $Q$ is maximum for $E$ = 0, then begins to decrease and takes minimal peaks at $E$ = $\Delta_{1}$. As $E$ increases, $Q$ becomes stable and continues linearly for all values of $\Delta_{1}$. In Fig. \ref{f7}b for $\phi_{i}$ = 1.5, we see that $Q$ decreases slightly. As $E$ increases, $Q$ varies linearly for all values of $\Delta_{1}$. It is  clear that the increase in  AB  flux erases the minimum peaks for $E$ = $\Delta_{1}$.

\begin{figure}[ht]
	\centering
	\includegraphics[scale=0.335]{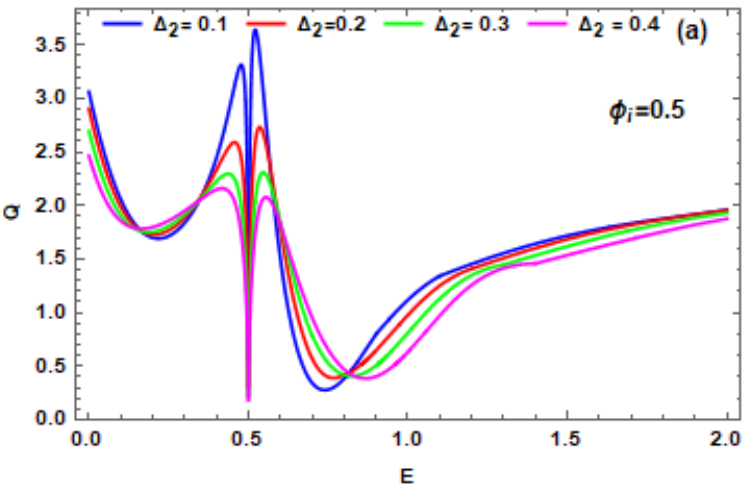}	
	\includegraphics[scale=0.335]{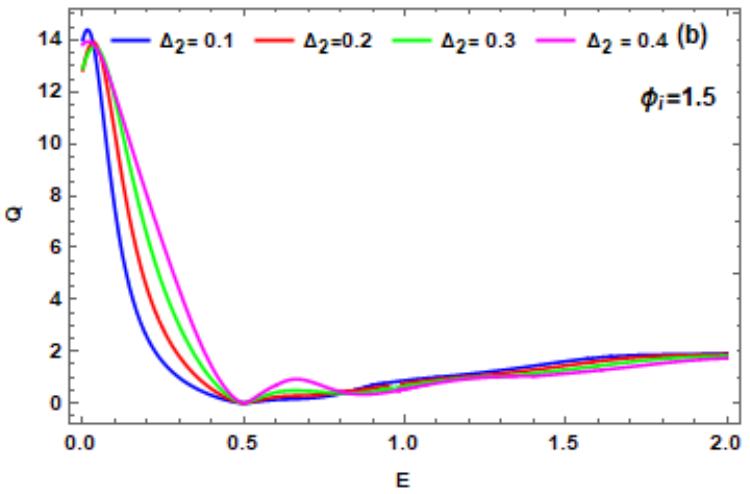}\\
	\caption{(color online) The same as Fig. \ref{f5} with inside gap values
		$\Delta_2 = 0.1, 0.2, 0.3, 0.4$,  $\Delta_1$=0.5, and $r_0$ =2.}
%
		\label{f8}
\end{figure}

We examine how $\Delta_{2}$ affects the scattering efficiency $Q$ as a function of the incident energy $E$ in Fig. \ref{f8}. For $\phi_{i}$=0.5, $Q$ is not zero at $E$=0 in Fig. \ref{f8}a. $Q$ first falls as $E$ rises, then rises again until it reaches $Q$=3.4. The value of $Q$ reaches a minimum peak around zero at $E$=$\Delta_{1}$=0.5. $Q$ falls in the energy range $\Delta_{1}<E<0.8$, and for all values of $\Delta_{2}$, $Q$ fluctuates linearly increasing as $E$ continues to increase. In Fig. \ref{f8}b, for $\phi_{i}$=1.5, $Q$ reaches its maximum value at $E$=0 and then decreases as $E$ increases until it reaches the critical point $E$=$\Delta_{1}$=0.5, after which it rises linearly.

\begin{figure}[ht]
	\centering
	\includegraphics[scale=0.335]{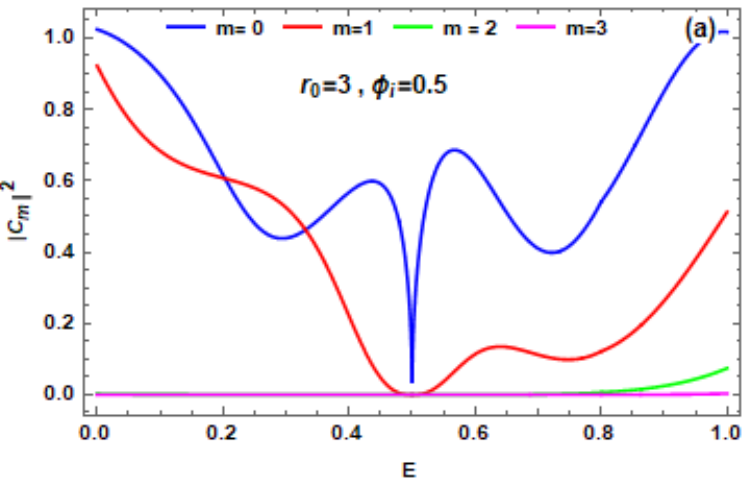}	
	\includegraphics[scale=0.335]{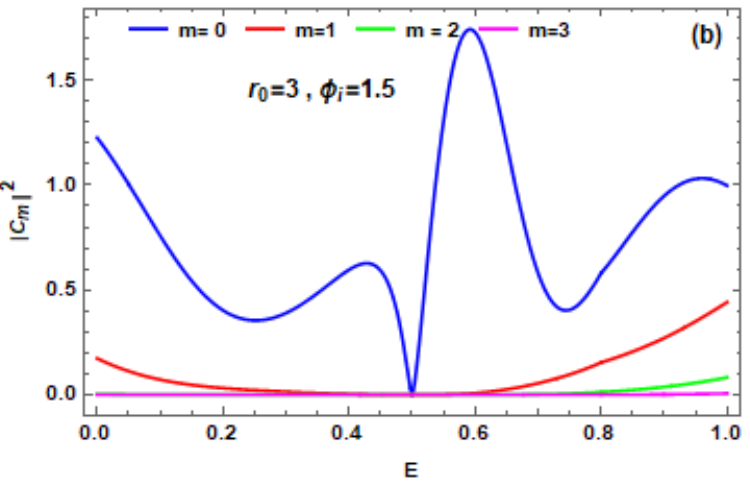}\\
	\includegraphics[scale=0.335]{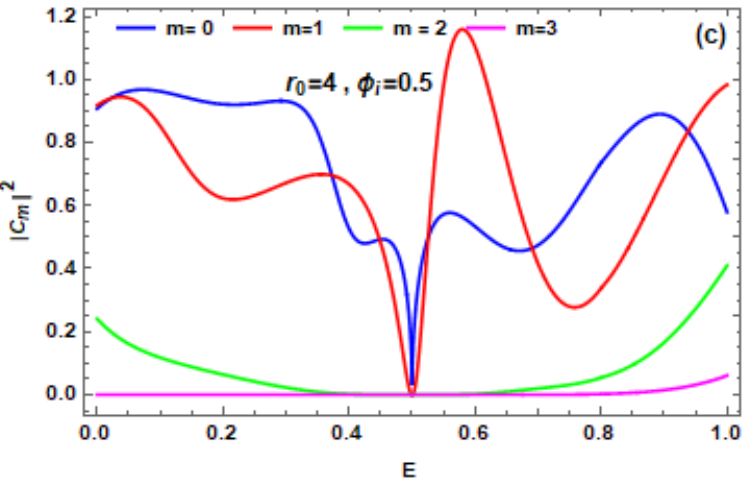}
	\includegraphics[scale=0.335]{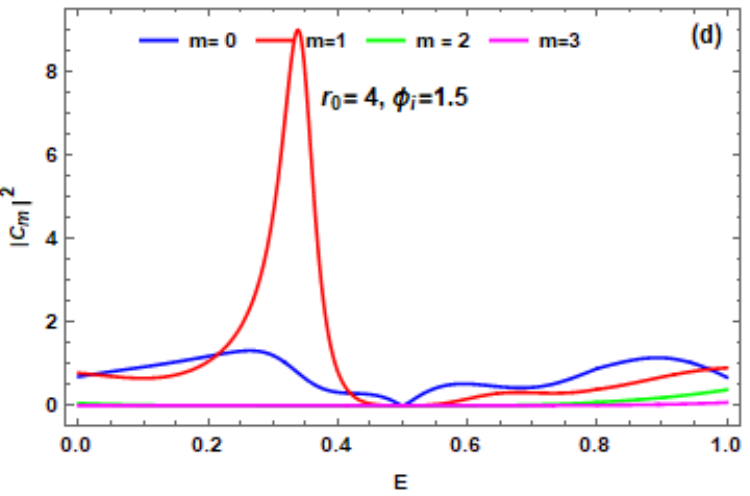}\\
	\includegraphics[scale=0.335]{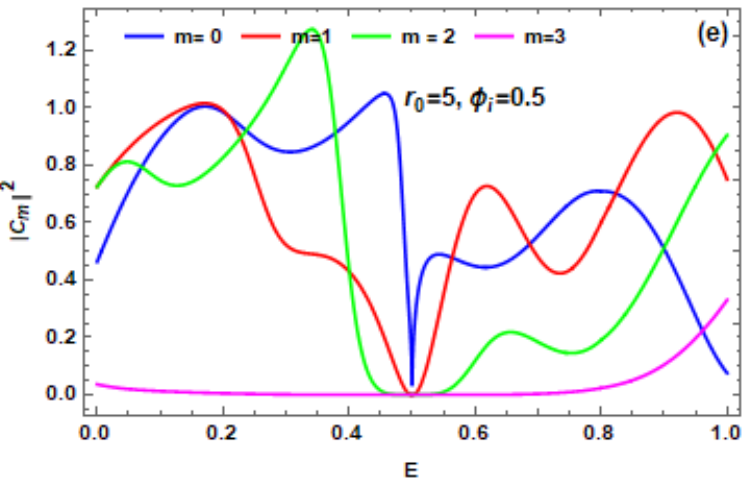}
	\includegraphics[scale=0.335]{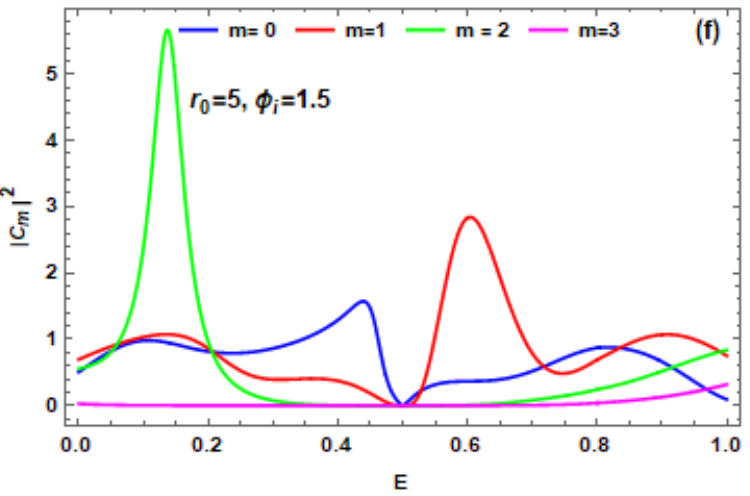}\\
	\includegraphics[scale=0.335]{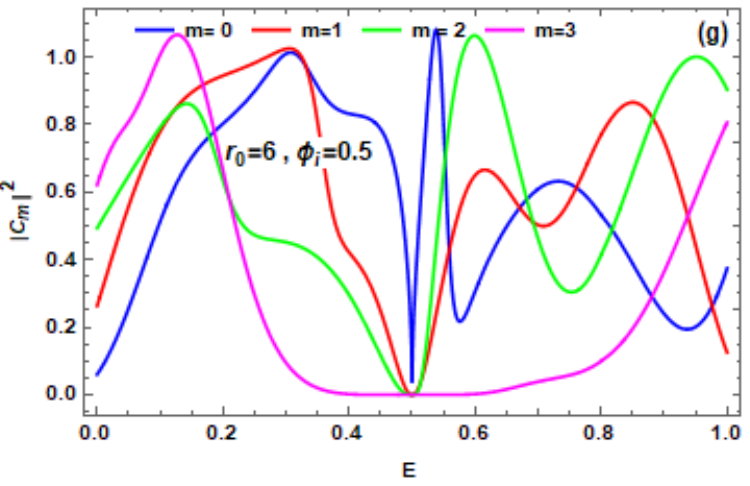}
	\includegraphics[scale=0.335]{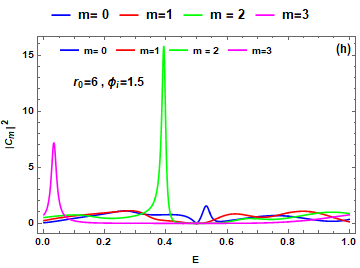}\\
	\caption {(color online) The square modulus of the scattering coefficients $|c_m(\phi_i)|^2$ as function of the incident energy $E$ for $m = 0, 1, 2, 3 $,  $V=1$, $\Delta_1$ = 0.5, and $\Delta_2$ = 0.2. (a,b): $r_0 = 3$, (c,d): $r_0 = 4$, (e,f):  $r_0 = 5$, (g,h):  $r_0 = 6$, with the first column $\phi_{i}$=0.5 and the second column $\phi_{i}$=1.5. }\label{f9}
\end{figure}

The effect of the size $r_0$ of the GQDs on the square model of the scattering coefficients $|c_{m}|^{2}$ as a function of the incident energy $E$ for angular momentum $m$=0,1,2,3 is illustrated in Fig. \ref{f9}. Indeed, if $\phi_{i}$=0.5 and $r_{0}=3.4$, then $|c_{m}|^{2}$ falls in the energy region $E<\Delta_{1}$, reaches a minimum value at $E=$$\Delta_{1}$, and shows oscillations for $E>$$\Delta_{1}$. 
For $r_0 = 5, 6$, $|c_m|^2$ first increases, then decreases, reaching a minimum at $E = \Delta_1$, after which it begins to oscillate again for $E > \Delta_1$. It is observed that the number of oscillations increases with $r_0$. These results are consistent with those reported in \cite{belokda}. For $\phi_i = 1.5$, $Q$ exhibits an almost linear variation, peaking at $E = \Delta_1$ for a given angular momentum $m$ and then becoming constant for all $m$ values when $E > \Delta_1$. This suggests that increasing the  AB flux $\phi_i$ leads to higher maximum values of $|c_m|^2$ and keeps it constant in the region where $E > \Delta_1$.

\begin{figure}[ht]
	\centering
	\includegraphics[scale=0.335]{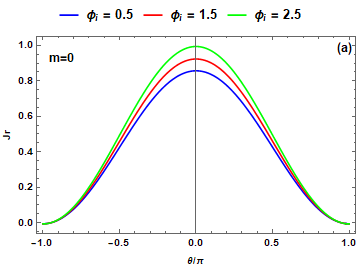}	
	\includegraphics[scale=0.335]{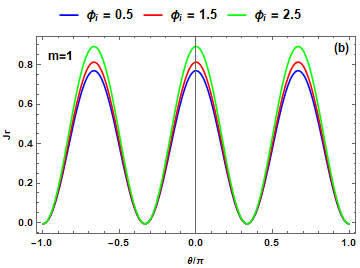}\\
	\includegraphics[scale=0.335]{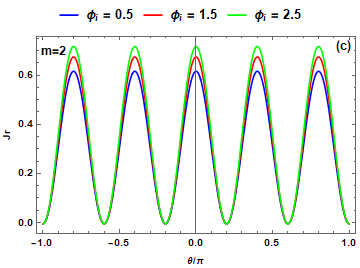}
	\includegraphics[scale=0.335]{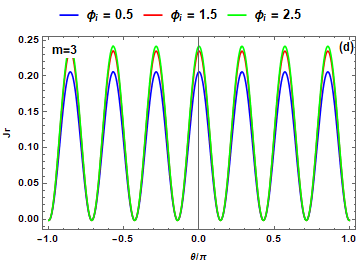}\\
	\caption  {(color online) The radial component of the reflected density current $J^{r}_{r}$ versus of the incident angle  $\theta$ for $V$ = 1, $\Delta_1$ = 0.1, $\Delta_2$ = 0.2, $E$= 0.6, $r_{0}$=7.8, and
	$\phi_{i}$=0.5, 1.5, 2.5, with	
	(a): $m$=0, (b): $m$=1, (c): $m$=2, and (d): $m$=3. }\label{f10}
\end{figure}

Fig. \ref{f10} shows the radial component of the reflected current density $J^{r}_{r}$ versus the incident angle $\theta$. We observe that $J^{r}_{r}$ varies periodically in an oscillatory manner with constant amplitude and depends on the  AB flux. In addition, the number of oscillations depends on $m$ such that for each value of $m$ there are $m+1$ oscillations of constant amplitude. Note that increasing the  AB flux increases the amplitude of $J^{r}_{r}$.

\section{Conclusion}

We have studied the scattering of Dirac electrons in GQDs subjected to a potential barrier $V$, two dual energy gaps ($\Delta_1$, $\Delta_2$), along with an external AB flux $\phi_i$. By applying suitable boundary conditions to the solutions of the Dirac equation, we were able to analyze the scattering phenomenon. Specifically, we calculated the radial component of the reflected current density $J^r_r$, the squared modulus of the scattering coefficient $|c_{m}|^{2}$, and the scattering efficiency $Q$.

 We have studied the scattering phenomenon by considering three energy regimes of incident electrons: $E<V_-$, $V_-<E<V_+$, and $E>V_+$. In the $E<V_-$ regime, we have shown that as the quantum dot radius $r_{0}$ approaches zero, $Q$ decreases and exhibits a peak at  AB flux values $\phi_{i}$=1.5 and 2.5. Lower values of $E$ correspond to larger amplitudes of $Q$. In the other regimes we observed a damped oscillatory behavior of $Q$ with increasing  AB flux. Increasing the external gap $\Delta_{1}$ led to higher values of $Q$.

We also looked at how the potential $V$ affected the scattering and discovered that when $V$ grows, $Q$ rises in the region where $E>\Delta_{1}$. In addition, peaks appeared at $E$=$\Delta_{1}$ when the  AB flux $\phi_{i}$ increased. The relationship between $Q$ and the size $r_{0}$ of the GQDs revealed maximum peaks for bigger sizes as $E$ approached $\Delta_{1}$. For $E<\Delta_{1}$, variable minimum peaks were observed at $\phi_{i}$=0.5, and $E>$$\Delta_{1}$, where there was linear behavior. When $\phi_{i}$=1.5, these peaks vanished.

We further analyzed the energy dependence of the squared modulus of the diffusion coefficients $|c_{m}|^{2}$, and found that at $E=0$ only the lowest coefficient is non-zero. As $E$ increases, other coefficients contribute, showing oscillatory behavior for $\phi_{i}$=0.5. However, for $\phi_{i}$=1.5, the oscillations disappear for larger size $r_{0}$ and are replaced by maximum peaks for specific values of $E$ and angular momentum $m$.

Finally, we have analyzed the radial component of the reflected current density, $J^r_r$, as a function of the incident angle $\theta$. It was observed that $J^r_r$ exhibits periodic oscillations of constant amplitude that are influenced by the  AB flux. Moreover, the number of oscillations depends on the angular momentum $m$, with each value of $m$ corresponding to $m+1$ oscillations of constant amplitude. In particular, increasing the AB  flux leads to an increase in the amplitude of $J^r_r$.

\end{document}